\documentclass[preprint,tightenlines,showpacs,preprintnumbers,amsmath,amssymb,aps]{revtex4}

\usepackage{graphicx}
\usepackage{bm}

\begin{document}
\draft
\title{Ferromagnetism of an all-carbon composite composed of
a carbon nanowire inside a single-walled carbon nanotube}

\author{Xiaoping Yang}
\affiliation{National Laboratory of Solid State Microstructures
and Department of Physics, Nanjing University, Nanjing 210093,
China,\\ and Department of Physics, Huainan Normal University,
Huainan, Anhui 232001, China }

\author{Jinming Dong}
\affiliation{National Laboratory of Solid State Microstructures
and Department of Physics, Nanjing University, Nanjing 210093,
China }

\begin{abstract}
Using the first-principles spin density functional approach, we
have studied magnetism of a new type of all-carbon nanomaterials,
i.e., the carbon nanowires inserted into the single-walled carbon
nanotubes. It is found that if the 1D carbon nanowire density is
not too higher, the ferromagnetic ground state will be more stable
than the antiferromagnetic one, which is caused by weak coupling
between the 1D carbon nanowire and the single-walled carbon
nanotube. Also, both dimerization of the carbon nanowire and
carbon vacancy on the tube-wall are found to enhance the magnetic
moment of the composite.\\
\end{abstract}

\pacs {73.22.-f, 75.75.+a, 61.46.+w, 71.15.Nc}
\date{16 July 2007}

\maketitle

All allotropes of carbon have long been considered as the most
promising candidates of the magnets [1]. Recent discovery of weak
ferromagnetism in pure carbon systems, such as carbon foam,
graphite, oxidized C$_{60}$ and polymerized rhombohedral C$_{60}$
[2-5], has stimulated renewed interests in new possible magnetic
materials made of only carbon. And also a lot of theoretical works
on the magnetism in all-carbon materials have been done [6-8].

Even so, the origin of ferromagnetism has not yet been clear. The
most important question is whether the spin polarization is
induced intrinsically or by magnetic impurities. If intrinsic,
what causes the unpaired spins, and how do they couple together
ferromagnetically? There are several theoretical models for the
magnetism in carbon-based compounds, which are nonbonding
localized states at the zigzag edge [6,9], carbon vacancies or
defects [4,7,10], and the trivalent carbon radicals [8].

The carbon nanotubes (CNTs) have attracted much attention since
their discovery [11] due to their remarkable electronic and
mechanical properties, and seem promising for future nanoscale
electronic devices [12]. A little change of the CNT structure,
e.g., the topological defect on tube wall, can greatly alter its
electronic properties, and the same holds when impurity atoms are
encapsulated into the CNT. Recently, X. Zhao {\it et al.}
discovered a new type of 1D carbon structures, carbon nanowires
(CNWs), formed by a linear carbon chain inserted into the
innermost tube ($\sim $ 7.0 {\AA} diameter) of a multiwalled CNT
[13]. On the other hand, however, the CNT magnetic properties have
been less studied [14], and so far, no experimental observations
show magnetism in CNTs.

In this letter, we investigate the electronic and magnetic
properties of the all-carbon composite, CNW@SWNT shown in Fig.
1(a), which is assumed to be composed of a CNW inserted into the
core of a single-walled CNT (SWNT). Our most striking result is to
find the flat-band ferromagnetism in various CNW@SWNTs at the
lower enough CNW densities, which is different from the previous
theoretical mechanisms [4,6,7-10].

Our theoretical calculations were performed with the total energy
plane-wave pseudopotential program CASTEP [15] in the local spin
density approximation (LSDA), in which the Ceperley-Alder form
exchange-correlation energy [16] was included. The ion-electron
interaction for the carbon atoms is modeled by the Vanderbilt
ultra-soft local pseudopotentials [17] with plane wave cut-off
energy of 240 eV. The CNW@SWNT is treated in a supercell geometry
[18] with intertube distance of 16 {\AA}, being larger enough to
prevent tube-tube interactions, and the supercell is optimized by
employing BFGS geometry optimization scheme [19].

We choose the armchair (5, 5) tube as the outer tube because its
diameter is the closest to that of the innermost tube observed in
the experiment [13]. The calculated band structures of the pure
(5, 5) tube, and the CNW@(5, 5) (model I in Fig. 1) are shown in
Fig. 2(a) and 2(b), respectively. The most interesting is
appearance of two flat-bands around $E_{F}$ with the first one
doubly degenerate, and their widths of about 0.05 and 0.4 eV,
respectively, which are clearly seen from Fig. 2(b), and are
absent in Fig. 2(a), demonstrating they are produced mainly by the
CNW. Since the CNW period of 4.89 {\AA} is larger than the
distance of 3.4 {\AA} between the graphite layers, the direct
coupling between the CNW atoms can be neglected. Thus, the narrow
flat-bands can be induced only by the weak coupling between the
CNW and the tube. As shown below, the flat-bands will cause the
higher density of states (DOS) at the $E_{F}$, and if the
interaction between the band electrons is introduced, the FM
ordering would be induced, which was proved exactly in the Hubbard
model under certain conditions [20].

The total DOS and partial DOS (PDOS) of the CNW@(5, 5) are shown
in Fig. 2c, d, e. It is seen from Fig. 2(c) that there exist
higher DOSs at and just below the $E_{F}$, which are almost
totally contributed by the CNW, identified by comparing Fig. 2(d)
with 2(e). More importantly, they are completely spin polarized,
making the model I composite a ferromagnetic (FM) metal with a
total magnetic moment of 2.44 $\mu _B $ in its one unit cell,
which is mainly contributed by the CNW atoms. In order to get its
real ground state, we then made the same LSDA calculation on a
doubled unit cell, and found the antiferromagnetic (AFM) state is
higher in energy than the FM one by 73.8 meV per Model I unit
cell, meaning the FM state is its ground state.

We also made the same LSDA calculations on other CNW@(5,5) with
different CNW periods, e.g., 1/2, 1 and 4/3 times that of the pure
(5,5) tube, respectively, and found that only if the CNW density
is not too higher, can the weak coupling between the CNW and (5,
5) tube induce the flat-bands at the $E_{F }$, producing thus
magnetic moments. For example, in the case of 1/2, the flat-band
will disappear, and so no ferromagnetism in its corresponding
composite of CNW@(5, 5). But, all of others have the flat-bands
and the magnetic moments.

It is well known that the Coulomb correlations will affect the
electronic states of weakly bonded carbon atom (here, e.g., the
CNW atoms), hence the magnetic property of the system. Therefore,
we have also made the LSDA+U calculation on the model I CNW@(5,5)
by using the projector augmented wave (PAW) potentials in the VASP
code [21,22]. In our calculations, the on-site Coulomb repulsion U
= 3.0 eV, and the exchange J = 0.9 eV have been taken only for the
CNW atoms. It is found that the FM state with a magnetic moment of
2.42 $\mu _B $ per Model I unit cell is lower in energy than the
AFM one by 53.5 meV per Model I unit cell, meaning the FM state is
always the ground state of the system.

The ferromagnetism found in the CNW@SWNTs reveals realization of
the flat-band ferromagnetism in the all-carbon composites with
only $s$ and $p$ orbitals. Analyses of the orbital density
distributions in the flat-band states show that they are composed
mainly of the $p$ orbitals of CNW atoms and less of the $\pi $
orbitals of tube-wall atoms. And they are localized around the CNW
atoms, but at the same time, extended along the tube direction
through the weak coupling between the CNW atoms and the SWNT,
which is clearly demonstrated by the small dispersion of the
flat-band shown in Fig. 2(b). Therefore, if the spins are
parallel, the repulsive interaction energy between the flat-band
electrons would be reduced on average, compared with the case of
anti-parallel spins, because of the Pauli exclusion, favoring
obviously more FM state than AFM one, which has also been proved
by the LSDA+U calculation mentioned above. In order to clarify
more certainly the mechanism for the ferromagnetism due to weak
coupling between the tube and the wire, a same LSDA+U calculation
on the {\bf isolated CNW} in Model I (in Fig. 1(a)) has been made,
demonstrating the FM state is higher in energy than the AFM one by
111 meV per CNW atom, meaning the {\bf AFM state} is the ground
state of the isolated CNW, in which the on-site Coulomb repulsion
of U = 3.0 eV, and the exchange J = 0.9 eV have been used.

What is effect of varous outer SWNTs with different diameters or
chiral symmetries on the magnetism of the composites? We select
two different CNW@SWNTs, such as CNW@(9,0) and CNW@(8,0), with
fixed CNW periods, which are the same as those of (9, 0) and (8,
0) tubes, respectively. By the same LSDA calculations we found
that both of them have magnetism with the magnetic moment of 2.41
$\mu _B $ and 2.24 $\mu _B $ per composite unit cell,
respectively, demonstrating no significant effects from a little
bit larger or smaller SWNT diameter and different chiral angles of
the outer tubes. So, the CNW@SWNTs composites may find wide-scale
technological applications.

Due to the well known Peierls instability, we found by the
first-principles LDA calculation that the dimerized phase of the
model I is lower in energy than the undimerized one by 81.6 meV
per model I unit cell, whose final optimization structure is shown
in Fig. 1 too, called as model II. In addition, we study also the
possible effect of the defect in real materials by introducing
carbon vacancy on the tube wall in model II, and its optimized
structure is called as model III in Fig. 1. For comparison, we
list in Table I all related LSDA results for the three models in
Fig. 1, from which, it is obvious that the FM behavior appears in
all the three composites. The model III has the largest magnetic
moment, which not only mainly comes from CNW atoms, but also is
partly contributed by the carbon atoms on the tube wall around the
vacancy. In addition, the charges in model III are transferred not
only from tube to the dimerized CNW, but also from some atoms to
the others on the same tube wall. The transferred charges between
the tube-wall atoms are mostly localized around the vacancy atom,
leading to appearance of the magnetic moment of - 0.04$\sim $0.9
$\mu _B $ per carbon atom on the tube wall, which is absent at all
in model I or II.

Finally, we show the spin densities [$n_\uparrow({\bf
r})-n_\downarrow({\bf r})$] (the up spin is the majority) of three
models in Fig. 3, from which it is clearly seen that in model I,
the spin density higher than 0.0068 electrons/{\AA}$^{3}$ exists
only at the CNW atoms, and in model II, it extends into the space
between two nearer CNW atoms. In model III, however, the same
magnitude spin density can also be found at the tube atoms nearby
the vacancy. So, both the CNW dimerization and the vacancy defect
on the tube wall can enhance the ferromagnetism of the composite.

In summary, we have studied several all-carbon CNW@SWNTs by the
first-principles LSDA and LSDA+U calculations, and identified
realization of the flat-band ferromagnetism in the composites if
the 1D CNW period is larger enough, which is caused by the weak
coupling between the 1D CNW and the outer tube. In addition, it is
found that the CNW dimerization and carbon vacancy on the tube
wall enhance the ferromagnetism of the composite. Our results
indicate that the CNW inside carbon nanotube is a promising new
candidate for the all-carbon nanometer-ferromagnets.

This work was supported by the Natural Science Foundation of China
under Grant No. A040108 and No. 90103038. X. P. Yang thanks Mr. J.
Zhou and H. M. Weng for their help in the numerical calculations.

\newpage

\begin{center}
\textbf{TABLE}
\end{center}

\begin{table}[htbp]
\textbf{Table I.} The LSDA-calculated magnetic moment of the FM
state in three models of the CNW@(5,5) shown in Fig. 1.
\begin{tabular}
{p{40pt}|p{180pt}|p{180pt}} \hline Model& \textit{magnetic moment
(}$\mu _B $\textit{) per unit cell of each Model}&
\textit{magnetic moment (}$\mu _B $\textit{) per CNW atom}\\
\hline  I&2.44& 2.42\\
\hline  II&5.04& 2.40\\
\hline III&5.93& 2.38\\
\hline
\end{tabular}
\label{tab2}
\end{table}

\newpage
\begin{center}
\textbf{Figure Captions}
\end{center}

Fig. 1 (Color online). Ball-and-stick model of the
different kinds of CNW@(5, 5). (a) The model I. Its CNW period is
two times larger than that of the pure (5, 5) tube. (b) The model
II, showing a dimerized CNW. (c) The model III, which is the same as
(b) except for existence of a carbon vacancy on tube-wall. The boxes
represent the unit cells.\\

Fig. 2 (Color online). Calculated band structures
for: (a) pure (5, 5) tube, (b) model I CNW@(5, 5). And the DOS of
the model I CNW@(5, 5): (c) Total DOS, (d) PDOS of the tube, and (e)
PDOS of the CNW. The blue and red represent the majority (up) and
minority (down) spin, respectively.\\

Fig. 3 (Color online). Spin density [$n_\uparrow({\bf
r})-n_\downarrow({\bf r})$] distributions in the composites shown in
Fig. 1. (a) model I. (b) model II. (c) model III. The cyan clouds
represent spin-densities exceeding 0.0068 electrons/{\AA}$^{3}$.


\begin{thebibliography}{99}
\bibitem{22} At the present time, the CASTEP code has not included the LSDA+U program, and
so we used the VASP package to do the LSDA+U calculations.
Generally the PAW potentials are more accurate than the ultra-soft
pseudopotentials.
\end{thebibliography}
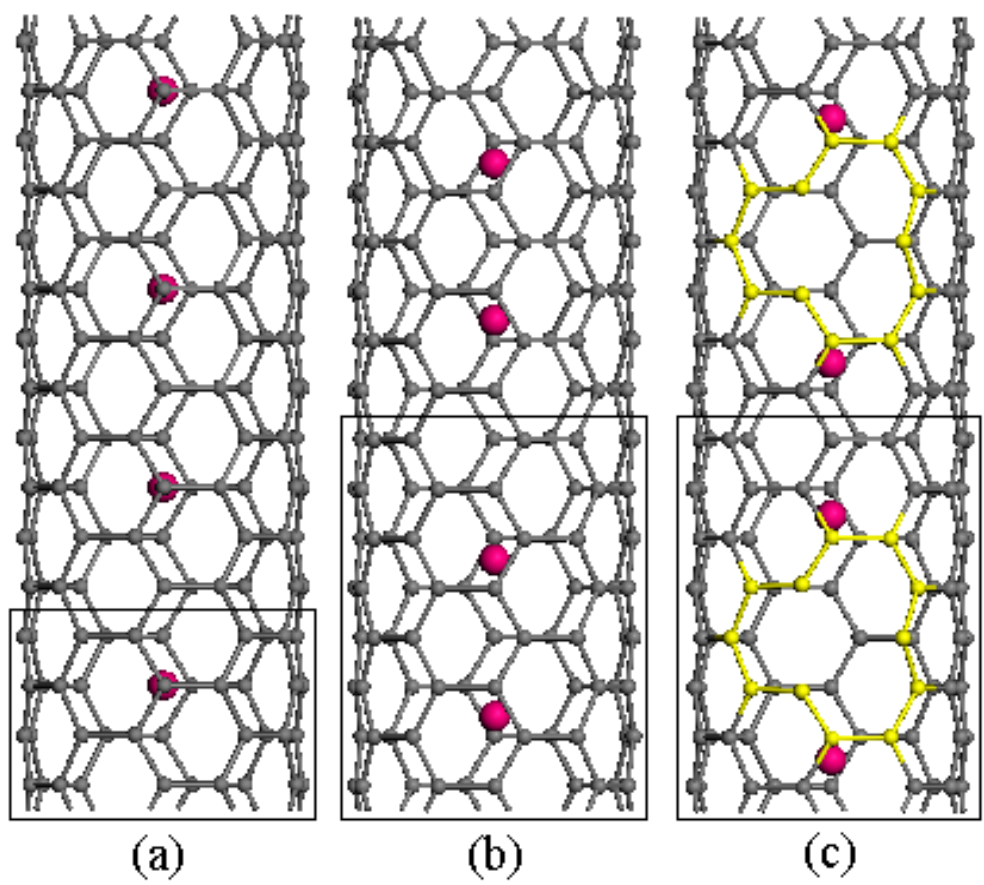
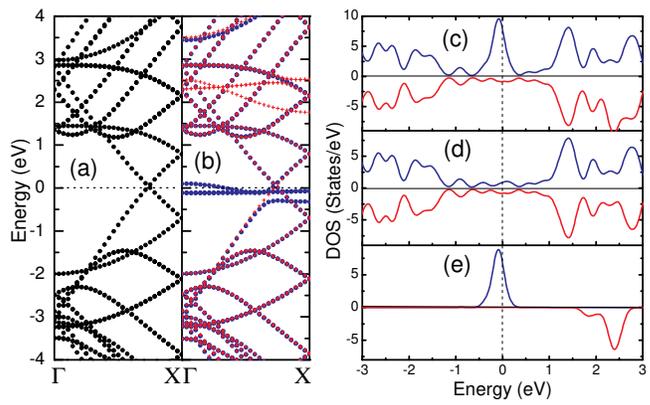
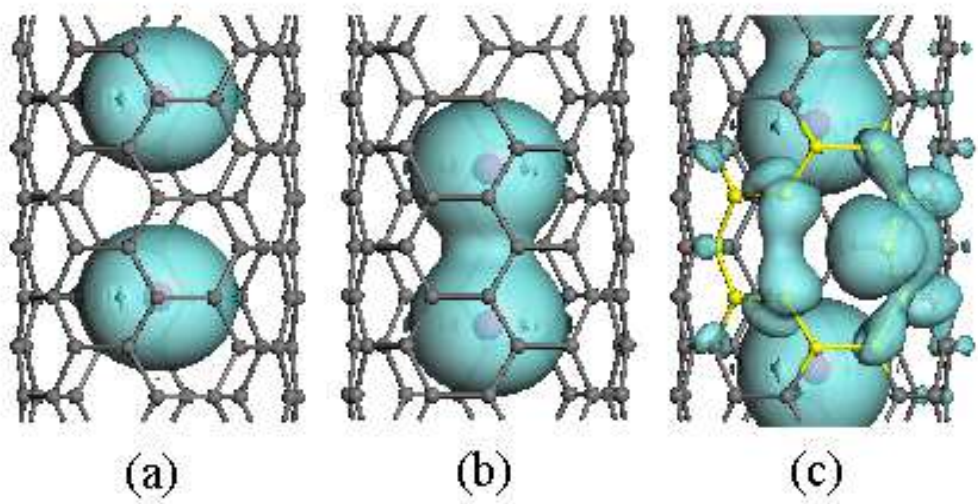
\end{document}